\def\be {\begin{equation}}
\def\ee {\end{equation}}
\def\bea {\begin{eqnarray}}
\def\eea {\end{eqnarray}}
\def\bc {\begin{center}}
\def\ec {\end{center}}
\def\bfg {\begin{figure}}
\def\efg {\end{figure}}
\def\bi {\begin{itemize}}
\def\ei {\end{itemize}}

\def\la {\label}
\def\le {\left}
\def\ri {\right}
\def\pa {\partial}
\def\fr {\frac}

\def\b  {\beta}

\def\C  {\Gamma}

\def\f {\phi}

\def\k  {\kappa}

\def\m  {\mu}

\def\o  {\omega}

\def\r  {\rho}

\def\s {\sigma}

\def\pa {\partial}

\documentclass[onecolumn,preprintnumbers,amsmath,amssymb]{revtex4}
\usepackage{graphicx}

\begin{document}

\title{Effect of Generalized Uncertainty Principle on Main-Sequence Stars and White Dwarfs}

\author{Mohamed Moussa}
\email{mohamed.ibrahim@fsc.bu.edu.eg ; moussa.km@gmail.com}
\affiliation{Faculty of Science, Physics Department, Benha University, Benha 13518, Egypt}

\begin{abstract}

This paper addresses the effect of generalized uncertainty principle, emerged by a different approaches of quantum gravity within Planck scale, on thermodynamic properties of photon, non-relativistic ideal gases and degenerate fermions. A modification in pressure, particle number and energy density are calculated. Astrophysical objects such as main sequence stars and white dwarfs are examined and discussed as an application. A modification in Lane-Emden equation due to a change in a polytropic relation caused by the presence of quantum gravity, is investigated. The applicable range of quantum gravity parameters is estimated. The bounds in the perturbed parameters are relatively large but it may be considered reasonable values in the astrophysical regime.

\end{abstract}

\maketitle

\section{Introduction}

Different approaches for the quantum gravity have been proposed in string theory and black hole physics to provide a set of predictions for a minimum measurable length and an essential modification of the Heisenberg uncertainty principle (GUP) \cite{1,2,3,4,5,6,7,8,9,10,11} and a modification in the fundamental commutation relation $[x_i, p_j]$. According to string theory, It has been found that  strings can not interact at distances smaller than their size, which yields  generalized uncertainty principle \cite{1}. From Black hole physics \cite{2,3}, the  uncertainty principle, $\Delta x \sim \hbar/\Delta p$, is modified at the Planck energy scale when the corresponding Schwarzschild radius is approximately equal to Compton wavelength at the Planck scale. Higher energies result in a further increase of the Schwarzschild radius, to yield $\Delta x \approx \ell_{Pl}^2\Delta p/\hbar$. The above approaches along with a combination of thought experiments and rigorous derivations suggest that the (GUP) holds at all scales and is represented by \cite{1,2,3,4,5,6,7,8,9,10,11}

\begin{eqnarray}
\Delta x_i \Delta p_i \geq \fr{\hbar}{2} [ 1 + \beta
\le((\Delta p)^2 + <p>^2 \ri)
+ 2\beta \le( \Delta p_i^2 + <p_i>^2\ri) ]~,
 \la{uncert1}
 \label{aa1}
\end{eqnarray}

where $p^2 = \sum\limits_{j}p_{j}p_{j}$, $\beta=\dfrac{\beta_0}{(M_{p}c)^2}=\b_0 \dfrac{\ell_{p}^2}{\hbar^2}$, $M_{p}$ is Planck mass, and $M_{p} c^2$ is Planck energy. It was shown in \cite{9}, that inequality (\ref{aa1}) is equivalent to the following modified Heisenberg algebra

\begin{eqnarray}
[x_i,p_j] = i \hbar ( \delta_{ij} + \beta \delta_{ij} p^2 +
2\beta p_i p_j )~.
\la{com1}
\label{aa2}
\end{eqnarray}
This form ensures, via the Jacobi identity, that $[x_i,x_j]=0=[p_i,p_j]$ \cite{10}. Different minimal length scale scenarios inspired by various approaches to the quantum gravity have been reviewed in \cite{12}. Apparently, this suggests a modification of the physical momentum \cite{9,10,11}

\begin{eqnarray}
p_i = p_{0i} \le( 1 +\beta
p_0^2\ri)~,
\label{mom1}
\label{aa3}
\end{eqnarray}

while $x_i = x_{0i}$ with $x_{0i}, p_{0j}$ satisfying the canonical commutation relations $ [x_{0i}, p_{0j}] = i \hbar~\delta_{ij}, $ such that \hbox{$p_{0i}=-i\hbar \partial/\partial{x_{0i}}$}, where $p$ is the momentum satisfying Eq.(\ref{aa2}). The upper bounds on the parameter $\beta$ has been derived in \cite{17,18} and it was found that it could predict an intermediate length scale between Planck scale and electroweak scale. It was suggested that these bounds can be measured using quantum optics techniques and gravitational wave techniques in \cite{19,20} which is considered as milestone in the quantum gravity phenomenology. It is noteworthy that the thermodynamical phenomenological implications of GUP in quantum gases are numerous, for example \cite{121,122}. In\cite{29} the quantum gravity influences on the statistic properties classical non-relativistic, ultra-relativistic and photon gases were addressed. Small corrections to the energy and entropy at low temperatures were found and some modifications to the equations of state is employed. The authors considered two configurations. The first is the white dwarf which is composed of a non-relativistic cold nuclei, the mass of the star was calculated. It is found that there is a quantum gravity correction depends on the number density of the star leads to an increase in the mass of the star and the quantum gravity tends to resist the collapse of the star. The other configuration is that the star is almost composed of ultra-relativistic particles. They found that the increase in the star energy leads to  monotonically blows up of Fermi pressure which resists the gravitational collapse.

We will use this form of GUP (\ref{aa1}) to investigate the implications of the quantum gravity on statistical properties of quantum gases. After redefining the new phase space we will put the partition function in a form that consistent with GUP and then rewriting the thermodynamic properties for quantum gases namely photons, non-relativistic ideal gases and fermions in degenerate state. Using these results we will follow the proposal which was investigated in reference \cite{21} in studding the stability of main-sequence stars and white dwarf. Also a simple method is used to study the effect of quantum gravity in a different stage of Stellar evolution and a modified mass-radius relation for the white dwarf is calculated and a boundaries in the quantum gravity perturbative terms will be determined.

\section{Statistical Mechanics and GUP}

In this section we investigate the GUP modification in the formalism of the grand-canonical ensemble. If we consider $N$ particles with energy states $E_i$. For each state there is $n_i$ particles in that state. The grand-canonical partition function is obtained from the sum over all of the states

\begin{eqnarray}
\mathcal{Z}= \sum_{n_i} \prod_i \le[e^{(\mu-E_i/K_BT)}\ri]^{n_i}
\label{1}
\end{eqnarray}

where $T$ is the temperature, $K_B$ is the Boltzmann constant and $\mu$ is the chemical potential. The connection to thermodynamics is obtained after introducing the partition function defined as \cite{31}

\begin{eqnarray}
\ln{\mathcal{Z}}  =\sum_i \ln\le[{1+a z e^{-E_i/{K_BT}}}\ri]
\label{2}
\end{eqnarray}

where $z=exp(\mu/K_BT)$, the parameter $a$ takes value depending on the considered particles. If they are fermions, $a$ takes $+1$ value, and if the considered particles are bosons, $a$ takes $-1$ value.

By considering large volume, the summation turns to be
$\sum_i\rightarrow \int \frac{d^3 x d^3 p}{(2 \pi \hbar)^3~ \le(1+\b p^2\ri)^3}$
where we are considering the invariant phase space in the existence of the GUP that has been derived in \cite{32}. We should put the grand canonical partition function for that systems in a form consistents with GUP frame work. The GUP can be considered in the phase space analysis by two equivalent pictures: (a) considering deformed commutation relations (i.e., deformed the measure of integration) simultaneously with the nondeformed Hamiltonian function or (b) calculating canonical variables on the GUP-corrected phase space which satisfy the standard commutative algebra (i.e., nondeformed standard measure of integration), but the Hamiltonian function now gets deformed. These two pictures are related to each other by the Darboux theorem in which it is quite possible to find canonical coordinates on the symplectic manifold which satisfy standard Heisenberg algebra \cite{23,22} . In this paper we are going to consider deformed measure of integration with nondeformed Hamiltonian function. Based on the new phase space the partition function

\begin{eqnarray}
\ln{\mathcal{Z}}=\dfrac{V}{(2 \pi \hbar)^3}\frac{g}{a } \int   \ln\le[{1+a z e^{-E/{k_BT}}}\ri]\frac{d^3 p}{ \le(1+\b p^2\ri)^3}
\label{4}
\end{eqnarray}

with $E=\le(p^2c^2+m^2c^4\ri)^{1/2}$. Pressure, number of particle and internal energy will be determined by the relations

\begin{equation}
P=K_BT\dfrac{\partial }{\partial V} \ln{\mathcal{Z}}~~~,~~~
n=K_BT\dfrac{\partial }{\partial \mu} \ln{\mathcal{Z}}\mid_{T,V}~~~,~~~
U=K_BT^2\dfrac{\partial }{\partial T} \ln{\mathcal{Z}}\mid_{z,V}
\label{5}
\end{equation}

We will expand all the terms that contains $\b $ and we will keep the terms that proportional to $\sim \b$ only. This approximation breaks down around maximum measurable energies such that $\b p^2\approx 1$. In this case we need an exact solution. But one can trust the perturbative solution where $\b p^2\ll 1$. Now we will use the modified partition fuction to calculate the thermodynamics of photons, non-relativistic ideal gases and degenerate fermions.

\subsection{Photon gas}

For a gas of photons $g=2$, $a=-1$ and $\m=0$. The energy density is given by, using Eq.(\ref{5}),

\begin{eqnarray}
 u=\dfrac{U}{V}=\dfrac{1}{\pi^2 \hbar^3}\int \dfrac{E}{e^{E/K_BT}-1} \dfrac{p^2 dp}{(1+\b p^2)^3}
\label{a1}
\end{eqnarray}

where $E=pc$. Use the approximation $1/(1+\b p^2)^3 \simeq 1-3\b p^2$, one gets

\begin{eqnarray}
\nonumber u=\dfrac{c}{\pi^2 \hbar^3}\int \dfrac{dp}{e^{cp/K_BT}-1}(p^3-3\b p^5)\\
=\dfrac{4\s}{c}T^4- \b \dfrac{8}{21}\dfrac{\pi^4 K_B^6}{\hbar^3c^5}T^6
\label{a2}
\end{eqnarray}

where $\s=\dfrac{\pi^2 k^4}{60 \hbar^3 c^2}$ is the Stefan-Boltzman constant. For $\b=0$ one can recover the usual Stefan-Boltzman law.

And the pressure of the system, Eq.(\ref{5}), is given by

\begin{eqnarray}
\nonumber P=\dfrac{c}{\pi^2\hbar^3}\int \dfrac{dp}{e^{cp/KT}-1}\le[\dfrac{1}{3}p^3-\dfrac{3}{5}\b p^5\ri]\\
=\dfrac{4\s}{3c}T^4- \b \dfrac{8}{105}\dfrac{\pi^4 K_B^6}{\hbar^3c^5}T^6
\label{a3}
\end{eqnarray}

It is clear that the modification in phase space decreases the available number of microstates, which holds the thermodynamic properties, and hence decreases the internal energy and pressure of the system. In order to obtain the pressure-energy density relation divide Eq.(\ref{a3}) by Eq.(\ref{a2})

\begin{eqnarray}
\dfrac{P}{u}= \dfrac{1}{3}+\b \dfrac{16}{21}\le(\dfrac{K_B\pi}{c}\ri)^2 T^2
\label{a4}
\end{eqnarray}

If we remove the quantum gravity effect we can recover the usual relation $u=3P$. In order to obtain the equation of state (EOS) $P=P(u)$ we first solve Eq.(\ref{a2}) for the temperature to get
\begin{equation}
T=\le(\dfrac{c}{4\s}\ri)^{1/4}u^{1/4}+\b \dfrac{10}{7} \le(\dfrac{\pi K_B}{c}\ri)^2 \le(\dfrac{c}{4\s}\ri)^{3/4} u^{3/4}
\label{a5}
\end{equation}

substitute Eq.(\ref{a5}) into Eq.(\ref{a4}) to get the first order correction to the EOS
\begin{equation}
P=\dfrac{u}{3}+\b \dfrac{16}{21} \le(\dfrac{15\pi^2\hbar^3}{c}\ri)^{1/2} u^{3/2}
\label{a6}
\end{equation}

Eq.(\ref{a6}) represents a modified equation of state due to generalized uncertainty principle.

\subsection{Non-Relativistic Ideal Gases}

Let us now consider the effect of quantum gravity on non-relativistic ideal gases. In order to do that the partition function for one particle with the modified phase space is given by

\begin{equation}
z_1=\dfrac{V}{2\pi^2\hbar^3}\int e^{-p^2/(2mK_BT)}\dfrac{p^2 dp}{(1+\b p^2)^3}
\label{C2}
\end{equation}

The total partition function is given by $Z=\frac{1}{N!}z_1^N$ by which we can drive the free energy for the system with the equation $F=K_B~T~ln Z$. Using Stirling's formula $ln N!=NlnN-N$ we can determine the pressure
\begin{equation}
P=-\dfrac{\pa F}{\pa V}\mid_{T,N}=\dfrac{N~K_B~T}{V}
\label{C3}
\end{equation}
So we recover the usual EOS for non-relativistic ideal gas $P=nK_BT$, where $n=N/V$ is the number density which shows that the quantum gravity does not play here. The same behavior is established in Ref.\cite{23}. Suffice calculation of the pressure at this point, which we need, and the other thermodynamic properties of the non-relativistic ideal gases can be obtained easily by solving the integration in Eq.(\ref{C2}) and then calculating the free energy.

\subsection{Degenerate Fermion Gas}

Fermions have a property that at most one can occupy each quantum state. At $T=0$ all states with energy less than Fermi energy level are occupied and all states with greater energy are empty. So the total energy of the system is the lowest possible consistent with the exclusion principle. The energy of the highest occupied state at absolute zero is the Fermi energy and corresponding momentum is referred to as the Fermi momentum $p_f$. In this case the distribution function goes to step function.  According to these assumptions and using Eq.(\ref{5}) the particle number density is given by

\begin{eqnarray}
\nonumber n=\dfrac{1}{\pi^2 \hbar^3} \int_{0}^{P_f} \dfrac{p^2 dp}{(1+\b p^2)^3}\\
=\dfrac{1}{\pi^2 \hbar^3} \le[\dfrac{1}{3}p_f^3-\b \dfrac{3}{5} p_f^5\ri]
\label{b}
\end{eqnarray}

ٍSolving this equation for Fermi momentum, one gets

\begin{eqnarray}
p_f=\le(3\pi^2\hbar^3n\ri)^{1/3}+\b \dfrac{3}{5}\le(3\pi^2\hbar^3n\ri)
\label{b1}
\end{eqnarray}

The second term ensures that the Fermi momentum is increased due to the presence of quantum gravity and hence Fermi energy. Also the energy density for fermions in degenerate state is given by

\begin{equation}
u=\dfrac{1}{\pi^2 \hbar^3} \int_{0}^{P_f}\le[p^2c^2+m^2c^4\ri]^{1/2} \dfrac{p^2 dp}{(1+\b p^2)^3}
\label{b2}
\end{equation}

use the substitutional

\begin{equation}
p=mc \sinh{x}
\label{b3}
\end{equation}

the energy density integration will be

\begin{eqnarray}
\nonumber u=\dfrac{m^4 c^5}{\pi^2 \hbar^3} \int_{0}^{x_f}\le[\sinh^2x+(1-3\b m^2 c^2)\sinh^4x-3\b m^2 c^2\sinh^6x\ri]dx \\
=\dfrac{m^4 c^5}{\pi^2 \hbar^3} \le[F_1(y)+(1-3\b m^2 c^2)F_2(y)-3\b m^2 c^2F_3(y)\ri]
\label{b4}
\end{eqnarray}

where
\begin{equation}
y=\dfrac{p_f}{mc}=\sinh{x_f}
\label{b5}
\end{equation}

with

\begin{equation}
F_1(y)=-\dfrac{1}{2}~ln\le[y+\sqrt{1+y^2}\ri]+\dfrac{1}{2}y\sqrt{1+y^2}
\label{b6}
\end{equation}
\begin{equation}
F_2(y)=\dfrac{3}{8}~ln\le[y+\sqrt{1+y^2}\ri]+\dfrac{1}{4}y\sqrt{1+y^2}\le[y^2-\dfrac{3}{2}\ri]
\label{b7}
\end{equation}
\begin{equation}
F_3(y)=-\dfrac{5}{16}~ln\le[y+\sqrt{1+y^2}\ri]+y\sqrt{1+y^2}
\le[\dfrac{5}{16}+y^2\le(\dfrac{y^2}{6}-\dfrac{5}{24}\ri)\ri]
\label{b8}
\end{equation}

by the same way, we can derive the pressure from Eq.(\ref{5}), after making use of the integration by parts and function expansion, one finds that

\begin{eqnarray}
\nonumber P=\dfrac{c^2}{\pi^2\hbar^3}\int_{0}^{p_f} \dfrac{dp}{\le[p^2c^2+m^2c^4\ri]^{1/2}}\le[\dfrac{1}{3}p^3-\dfrac{3}{5}\b p^5\ri]\\
=\dfrac{m^4 c^5}{\pi^2 \hbar^3} \le[\dfrac{1}{3}F_2(y)-\dfrac{3}{5}\b m^2 c^2F_3(y)\ri]
\label{b9}
\end{eqnarray}

Eq.(\ref{b9}) represents the ground state pressure. There is a considerable change due to the presence of the function $F_3(y)$ and $y$ itself depends on modified Fermi momentum. We should notice that the value of $y$ is increased due to the increasing in Fermi momentum, this leads to a considerable growing in the pressure Eq.(\ref{b9}). The change in pressure of the radiation and degenerate fermions  will affect the polytropic relations for astro objects. In the next section the modified pressure effect in the some stars stability and some other applications will be investigated exactly.

\section{Applications to Astrophysical Objects}

\subsection{Main-Sequence Stars}
Most main-sequence stars have a mass in the range about $10$ to $50$ solar mass. They are composed by a mixture of non-relativistic gas and radiation that are stood in hydrostatic equilibrium by gravity \cite{40}. The total pressure is $P_c=P_g+P_r$ where $P_c$,$P_g$ and $P_r$ are center star, gas and radiation pressure, respectively. We can write $\b_c=P_g/P_s$ and $1-\b_c=P_r/P_c$. By definition $(1-\b_c)$ and $\b_c$ are the fractional contribution of radiation and gas to the central pressure of the star and both are less than one, then

\begin{equation}
\dfrac{P_g}{\b_c}=\dfrac{P_r}{1-\b_c}
\label{E}
\end{equation}

The classical gas pressure can be expressed in the form, using Eq.(\ref{C3}),

\begin{equation}
P_g=\dfrac{\r_c~K_B~T}{\bar{m}}
\label{1E}
\end{equation}

We used the relation $n=\frac{\r_c}{\bar{m}}$, $\r_c$ is the central density of the star and $\bar{m}$ is the average mass of the gas particle that construct the star defined as $\bar{m}=u_Nm_N$ where $u_N$ is the mean molecular weight and $m_N$ is the nucleon mass. Use Eqs.(\ref{a3},\ref{1E}) into Eq.(\ref{E}), one finds that

\begin{equation}
\dfrac{\r_c K_B}{\b_c~\bar{m}} T_c=\dfrac{4\s}{3c(1-\b_c)}T_c^4- \dfrac{8}{105}\b \dfrac{\pi^4 K_B^6}{\hbar^3 c^5(1-\b_c)}T_c^6
\label{E1}
\end{equation}

Where $T_c$ is the central temperature of the star. We can fine the temperature by solving the above equation

\begin{equation}
T_c=\le(\dfrac{1-\b_c}{\b_c \bar{m}}\dfrac{3cK_B}{4\s }\ri)^{1/3}\r_c^{1/3}+\b \dfrac{1}{70}\dfrac{(1-\b_c)}{\b_c \bar{m}}\dfrac{\pi^4 K_B^7}{\hbar^3 c^3\s^2 }~\r_c
\label{E2}
\end{equation}

use this expression to calculate the star pressure from the relation $P_s=\frac{P_g}{\b_c}=\frac{\r_c K_B}{\b_c \bar{m}}T_c$ one gets

\begin{eqnarray}
\nonumber P_c=K_1 \r_c^{4/3}+\b K_2 \r_c^2 \\
=P_{sUP}+p_{sGUP}
\label{E3}
\end{eqnarray}

\begin{equation}
K_1=\dfrac{(1-\b_c)^{1/3}}{(\b_c)^{4/3}}\le(\dfrac{3cK_B^4}{4\s \bar{m}^4}\ri)^{1/3} ~~~~,~~~~K_2=\dfrac{1}{70}\dfrac{(1-\b_c)}{(\b_c)^2}\dfrac{\pi^4 K_B^8}{\hbar^3 c^3\s^2\bar{m}^2}
\label{E4}
\end{equation}

The first part of Eq. (\ref{E3}) is the ordinary polytropic relation with a a polytropic exponent $\C=\frac{4}{3}$ (polytropic index $n_p=3$) \cite{401} perturbed by a polytrope $\C=2$ (polytropic index $n_p=1$) where $n_p=\frac{1}{\C-1}$.

The best example may be used to estimate GUP perturbation is our sun. If we assume that $\b_c$ is constant overall the stars and its chemical composition is not changed so $u_N$ is constant. For our sun $1-\b_c=10^{-3}$ and $u_N=0.829$ \cite{41} and central density $\r_c=1.53\times10^5~kg/m^3$, then

\begin{eqnarray}
\dfrac{p_{sGUP}}{P_{sUP}}=1.8\times10^{-49} \b_0
\label{E9}
\end{eqnarray}

The perturbed pressure should not be exceeds the original one such that $\frac{p_{sGUP}}{P_{sUP}}\ll 1$, so this equation suggests that the upper value of $\b_0 $ should be $\b_0< 10^{48}$. This bound is far weaker than that set by electroweak measurements.

According to Eq.(\ref{E3}) the quantum gravity modifies the polytropic relation by adding a new term proportional to $\r^2$. This suggests a new modification in Lane-Emden equation \cite{43}. The modified Lane-Emden equation is established in appendix A. Studding of modified Lane-Emden equation and its solution is not the scope of this paper so results that are investigated by the this equation will be postpone to another research.
\\\\
\textbf{Main-sequence stars stability}
\\

In order to discuss the stability concept, the energy of a Newtonian polytrope systems with pressure $P$ can be expressed in terms of the internal energy and gravitational potential such that

\begin{equation}
E=k_1PV-k_2\dfrac{GM^2}{R}
\label{E5}
\end{equation}

where $k_1$ and $k_2$ are constants, $G$ is the gravitational constant and $M$ and $R$ is the mass and radius of the star. At hight densities it is important to take into account the general relativistic effects in star stability. To deal with it a general relativistic correction term should be added to Eq.(\ref{E5}), in lowest order approximation, namely $E_{corr}=-k_3 \le(\frac{G}{c}\ri)^2  M^{7/3}\r_c^{2/3}$  where $k_3$ is constant depends in the actual distribution of matter. According to \cite{41}, for $n_P=3$ polytrope, the condition of stability is $\C>\frac{4}{3}+2.25\frac{GM}{c^2R}$. We can see that addition of a relativistic correction contracts the region of stability or increase the critical value of $\C$. In our case the $\frac{GM}{c^2R}\sim~10^{-6}$ so this correction can be neglected. So we expect that the new term in a polytropic equation (\ref{E3}) with $n_p=1$, which has a positive contribution, will not affect the problem of stability since $\C > 2$.

Using Eq.(\ref{E3}) into Eq.(\ref{E5}) the energy of main-sequence stars will be

\begin{equation}
E=C_1M\r_c^{1/3}+\b C_2M\r_c-k_2M^{5/3}\r_c^{1/3}
\label{E6}
\end{equation}

where $C_1$ and $C_2$ are constants. The mass $M$ can be obtained by using the equilibrium condition $\frac{\pa E}{\pa \r_c}=0$, the result is

\begin{equation}
M=\le[\dfrac{3}{k_3}\le(\dfrac{1}{3}C_1+\b C_2\r_c^{2/3}\ri)\ri]^{3/2}
\label{E7}
\end{equation}

then

\begin{equation}
\dfrac{d~\ln{M}}{d~\ln{\r_c}}=\b \dfrac{3C_2}{C_1}\r_c^{2/3}
\label{E8}
\end{equation}

Eq.(\ref{E8}) ensure that $\frac{d~\ln{M}}{d~\ln{\r_c}}>0$ which is the stability condition \cite{41} for main sequence object. These results show that the generalization process for the uncertainty principle does not change the stability of the main sequence objects.
\\\\
\textbf{Minimum and maximum masses for main-sequence stars}
\\

In this part we will discuss the evolution stage of main-sequence stars in the presence of quantum gravity to derive the minimum and maximum masses of these objects. Initially, in the stage of star forming, the star must has mass enough to generate a central temperature which is high enough for thermonuclear fusion. At this stage the central pressure comes by the electrons and ions which are forming an ideal classical gas, their pressure can be expressed as in Eq.(\ref{1E}). The star will be made if this pressure is close to the pressure needed to support the system. It is known that the upper limit relation between the pressure and density at the center of the star is given by \cite{70}

\begin{equation}\label{1x}
P_c \leq \le(\dfrac{\pi}{6}\ri)^{1/3} G~ \r_c^{4/3}~M^{2/3}
\end{equation}

This relation is valid for any homogeneous star in which the mass is concentrated towards the center. Equating Eq.(\ref{1E}) and Eq.(\ref{1x}) one can get the initial central temperature in this stage

\begin{equation}\label{2x}
K_BT_c \leq \le(\dfrac{\pi}{6}\ri)^{1/3}~ \bar{m}~G \r_c^{1/3}~M^{2/3}
\end{equation}

We can see that the quantum gravity does not appeared in the picture. As shown in this equation the central temperature will rise as the central density increasing with star contraction. The contracting temperature continue to rise until the electrons at the center become in a fully degenerate, in a lowest energy states in accordance with exclusion principle. At this stage, at the center the pressure comes from the electrons that are in degenerate state plus the pressure from the classical ions. Quantum gravity will not change the classical ions pressure, according to Eq.(\ref{C3}), but a non-relativistic degenerate electrons pressure will change according to Eq.(\ref{b9}). In non-relativistic limit, where $y \ll 1$, the degenerate pressure, Eq.(\ref{b9}), will be

\begin{equation}\label{33x}
P_0^{non-relat}=\dfrac{1}{15}\dfrac{m_e^4~c^4}{\pi^2~\hbar^3}y^5-\dfrac{3}{35}\dfrac{m_e^6~c^7}{\pi^2~\hbar^3}y^7
\end{equation}

Use Eqs.(\ref{b1}, \ref{b5}) in above equation and add the result to Eq.(\ref{C3}), finally we can write the central pressure at this stage as

\begin{equation}\label{3x}
P_c=\dfrac{a}{5~m_e}~n_e^{5/3}+~\b~\dfrac{12~a^2}{35~m_e}~n_e^{7/3}+n_i~K_B~T_c
\end{equation}

where $a=(3^{1/3}\pi^{2/3}\hbar)^2$. The number density of the electrons $n_e$ and ions $n_i$ can be expressed in terms of central density, simply $n_e=n_i=\r_c/m_H$, where $m_H$ is the mass of hydrogen atom. A hydrostatic pressure is achieved if this pressure equals to the pressure needed to support the mass Eq.(\ref{1x}). This leads to the a central temperature to be

\begin{equation}\label{4x}
K_BT_c\leq \le(\dfrac{\pi}{6}\ri)^{1/3}G~m_N~M^{2/3}~\r_c^{1/3}-\dfrac{a}{5~m_e}~\le(\dfrac{\r_c}{m_H}\ri)^{2/3}   -\b~\dfrac{12~a^2}{35~m_e}~\le(\dfrac{\r_c}{m_H}\ri)^{4/3}
\end{equation}

The first term is associated with classical ions and second and third terms with degenerate electrons. The last two terms become important and are increasing with central density until the temperature will cease to rise as the mass contracts. From this equation we can find the maximum value of temperature

\begin{equation}\label{5x}
\le| K_BT_c \ri|_{max} \leq z~M^{4/3}\le[1-\dfrac{60}{7}~\b~m_e~z~M^{4/3}\ri]
\end{equation}

\begin{equation}\
z=\dfrac{5}{4}\le(\dfrac{\pi}{6}\ri)^{2/3}\dfrac{m_e}{a}~m_H^{8/3}~G^2
\end{equation}

Equation (\ref{5x}) prove that the maximum central temperature decreases due to quantum gravity. The contracting mass achieves stardom if that maximum central temperature reaches ignition temperature for the thermonuclear fusion of hydrogen. Denotes this ignition temperature by $T_{ign}$ then the minimum mass for the star will be given by

\begin{equation}\label{6x}
M_{min} \geq \le(\dfrac{K_BT_{ign}}{z}\ri)^{3/4}\le[1+\dfrac{45}{7}~\b~m_e~K_BT_{ign}\ri]
\end{equation}

This relation shows the quantum gravity effect increases the minimum mass of the main sequence stars. The approximation has a meaning if the second term in Eq.(\ref{6x}) is less than unity. If we take the ignition temperature for hydrogen to be about $1.5\times 10^6~K$, this suggests that the upper value of $\b_0<10^{47}$.
\\\\
The situation is changed if the radiation pressure takes a part in this stage of revolution. The star will be disrupted if radiation becomes the dominant source for the internal pressure. This property will set a limit to the mass of a main sequence star. The pressure at the center of hot massive stars is due to electrons and ions (as a classical gases) and radiation, the pressure that is produced by these species is derived in Eq.(\ref{E3}). Compere this with the pressure needed to support the star, Eq.(\ref{1x}), to get the maximum mass, one finds that

\begin{equation}\label{7x}
\le(\dfrac{\pi}{6}\ri)^{1/3}G~M_{max}^{2/3}\geq K_1~+\b~K_2~\r_c^{2/3}
\end{equation}

which means that the the maximum mass is increased due the the presence of quantum gravity. this increase depends on the central density. Again the second term in the right hand side of inequality (\ref{7x}) should be less than the first, this suggests that $\b_0<10^{48}$.

\subsection{White Dwarfs}

The white dwarfs  consist of Helium atoms that is almost in an ionization state. The temperature of the star is about $10^7 K$, so the electrons are in a relativistic energy regime and can be considered in a complete degenerate state and the Helium nuclei do not play a part, in first order approximation, in the dynamic of the system. Now we will determine the effect of GUP on white dwarf radius according to pressure modification expressed in Eq.(\ref{b9}). If the number of Helium nuclei is $N/2$ then there are $N$ electrons. Suppose that $M=2Nm_p$ is the mass of the star, so the particle density of the electron is

\begin{eqnarray}
n_s=\dfrac{3 M}{8\pi m_p}\dfrac{1}{R^3}
\label{e1}
\end{eqnarray}

Use Eqs.(\ref{b1},\ref{e1}) into Eq.(\ref{b5}), one gets

\begin{eqnarray}
y=\dfrac{\overline{M}^{1/3}}{\overline{R}}+ \dfrac{3}{5}~\b~m_e^2~c^2\dfrac{\overline{M}}{\overline{R}^3}
\label{e2}
\end{eqnarray}

\begin{eqnarray}
\overline{M}
=\dfrac{9\pi}{8}\dfrac{M}{m_p}~~~~,~~~~\overline{R}=\dfrac{m_e c}{\hbar} R
\label{e3}
\end{eqnarray}

where $R$ is the radius of the star and $m_p$ is the proton mass.
The equilibrium configuration of white dwarf is caused by the pressure of the degenerate electrons against Newtonian gravitational collapse which is due to the nuclei, which means that

\begin{eqnarray}
 P_0=\dfrac{\k}{4~ \pi} \dfrac{G~M^2}{R^4}=K'~\dfrac{\overline{M}^2}{\overline{R}^4}~~~~~,~~~~~K'=\dfrac{\k~ G}{4~\pi}\le(\dfrac{8~m_p}{9\pi}\ri)^2\le(\dfrac{m_e~c}{\hbar}\ri)^4
\label{e4}
\end{eqnarray}

We have two extreme cases as follows.

(i) When the electron gas is in a low density, non-relativistic dynamics may be used
such that$y\ll 1$, Eq.(\ref{b9}) will leads to.

\begin{eqnarray}
 P_0=\dfrac{4}{5}K~y^5-\dfrac{36}{35}~\b~m_e^2~c^2~K~y^7~~~~~,~~~~~ K=\dfrac{m_e~c^2}{12\pi^2}\le(\dfrac{m_e~c}{\hbar}\ri)^3
 \label{e5}
\end{eqnarray}

Equating Eq.(\ref{e4}) and Eq.(\ref{e5}), we got the equation

\begin{eqnarray}
\overline{R}^3-\dfrac{4}{5}~\dfrac{K}{K'}~\overline{M}^{-1/3}~\overline{R}^2-~\dfrac{48}{35}~\b~m_e^2~c^2~\dfrac{K'}{K}~\overline{M}^{1/3}=0
 \label{e6}
\end{eqnarray}

The solution of that equation is

\begin{eqnarray}
\overline{R}\simeq \dfrac{4}{5}~\dfrac{K}{K'}~\overline{M}^{-1/3}+\dfrac{15}{7}~\b~m_e^2~c^2~\dfrac{K'}{K}~\overline{M}
 \label{e7}
\end{eqnarray}

This equation shows that the radius of the star is increasing, the elevation depends in the mass of the star. The used approximation is true if the second term in the right hand side of Eq.(\ref{e7}) is much less than the first one, this will but a limit in the upper value of $\b_0$ such that it should be $\b_0<10^{44}$, we used a white dwarf mass $M=0.41~M_{\odot}$ \cite{44}.

(ii) when the electron gas is in a high density that relativistic effect comes to play strongly such that $y \gg 1$. We can write

\begin{eqnarray}
P_0=K \le(y^4-y^2\ri)-\b~\dfrac{18}{5}~ K~ m_e^2~c^2~\le(\dfrac{y^6}{3}-\dfrac{y^4}{4}\ri)
\label{e10}
\end{eqnarray}

Using Eq.(\ref{e10}) into Eq.(\ref{e4}) and simplify, then

\begin{eqnarray}
\overline{R}^4+\le[\dfrac{K'}{K}~\overline{M}^{4/3}-\le(1-\dfrac{3}{10}~\b~ m_e^2~c^2~\ri)\overline{M}^{2/3}\ri]~\overline{R}^2-
\dfrac{6}{5}~\b~ m_e^2~c^2~\overline{M}^{4/3}=0
\label{e11}
\end{eqnarray}

The solution of that equation, keeping only the terms that proportional to $\sim\b$, is

\begin{eqnarray}
\overline{R}\simeq \overline{R}_0\le[1+\b~\dfrac{3}{10}\dfrac{m_e^2~c^2}{\overline{R}_0^2}
\le(2~\dfrac{\overline{M}^{4/3}}{\overline{R}_0^2}-\dfrac{1}{2}\overline{M}^{2/3}\ri)\ri]
\label{e13}
\end{eqnarray}

where

\begin{eqnarray}
\overline{R}_0^2=\overline{M}^{2/3}-\dfrac{K'}{K}~\overline{M}^{4/3}
\label{e12}
\end{eqnarray}

Use Eq.(\ref{e3}) into Eq.(\ref{e12}) and solving for $R$ and let $R\rightarrow R_{Ch}$,one gets

\begin{eqnarray}
R_{Ch}=\dfrac{(9~\pi)^{1/3}}{2}\dfrac{\hbar}{m_ec}
\le(\dfrac{M}{m_p}\ri)^{1/3}\le[1-\le(\dfrac{M}{M_{Ch}}\ri)^{2/3}\ri]^{1/2}~~~~~,~~~~~M_{Ch}=\dfrac{9~(3~\pi)^{1/2}}{64~\k}\le(\dfrac{\hbar~ c}{G~m_p^2}\ri)^{3/2}~m_p
\label{e14}
\end{eqnarray}

$R_{Ch}$ is the unmodified white dwarf radius or Chandrasekhar radius limit and $M_{Ch}$ is the Chandrasekhar limit mass, of order $1.44M_\odot$. First part of Eq.(\ref{e13}) is the Chandrasekhar radius and the second term is the radius correction due to GUP affect. These analysis show that the increasing in star radius depends on $\le[1-\le(\frac{M}{M_{Ch}}\ri)^{2/3}\ri]^{-1}$ which formally goes to infinity as $M\rightarrow M_{Ch}$. It is clear that the quantum gravity effect become significant when mass of the white dwarf gets very close to the Chandrasekhar limit in a high density stars. Equations (\ref{b9}), (\ref{e7}) and (\ref{e13}) attract attention to a noticeable point is that the growing in star radius ensures that quantum gravity resists gravitational collapse by increasing the electrons degeneracy pressure. These results consistence with that established in Ref.\cite{29}. Our approximation will have a meaning if the second term in the right hand side in Eq.({\ref{e13}}) is less than one, this put an upper limit for $\b_0$ such that $\b_0<10^{43}$. We have used $M=0.59~M_{\odot}$ \cite{44}.

We should also notice that in the limit $y\gg 1$, in ultra-relativistic regime, the function $F_3(y)$ behaves like $\sim y^6$ which almost goes like $\r^2$. This result is consistent with polytrope behavior in which $\Gamma =2$, i.e., corresponding to stability range of the star. So we can conclude that this approach of quantum gravity does not produce any instability to white dwarfs.

\section{Discussion}

Various approaches to quantum gravity such as string theory, black hole physics, and doubly special relativity predict a considerable modification in Heisenberg uncertainty principle to be a generalized uncertainty principle. This modification leads to a change in the energy-momentum dispersion relation and in the physical phase space. In this paper we studied the modification due to a generalized uncertainty principle, Eq.(\ref{aa1}), in the statistical and thermodynamic dynamic properties for photons, non-relativistic ideal gases and degenerate fermions. There is a considerable decrease in the number of accessible microstates in quantum gravity background. This approach leads to a decrease in the internal energy and pressure of photon gas. Pressure and internal energy of fermions in degenerate state are increased and pressure of non-relativistic ideal gases is not changed. These results are used in studying the stability of the main-sequence stars and white dwarfs. It is found that this approach of quantum gravity does not produce any changes in stars stability.

We follow a very simple way to study the effect of quantum gravity in a different stage of stellar evolution. It is found that there is no change in pressure or temperature in the initial stage where the pressure is coming from classical ions. In the second stage , after stardom contracting, where degenerate electrons domain the central pressure, the quantum gravity leads to a decrease in the central maximum temperature and an increasing in the minimum and maximum masses of the main-sequence stars. This results shows that the quantum gravity has an effect in the building process of the universe and it well have a more strong effect in a highly dense objects.

A mass-radius relation for white dwarf is also investigated. The results show that there is an elevation in the star radius that is proportional to the mass of the star. For the case of high dense stars the elevation in the radius goes to infinity as the mass of the star reaches Chandrasekhar limit. This means that quantum gravity resists gravitational collapse. This results goes in parallel with that found in \cite{29}. Unfortunately, quantum gravity behaves in a different way from the prediction of the usual model for white dwarf. The current observation indicates that white dwarfs have smaller radii than theoretical predictions \cite{44,45,46}. This is not consistent with our results. But this result may be reasonable if we add this perturbed term, which comes from quantum gravity, to the other perturbed terms that are coming from other physical aspects such as Coulomb corrections, lattice energy, correction due rotation and magnetic field (which may be much larger than the correction due to GUP) and more realistic density distribution for electronic gas. The sum of all these terms may be consistent with the current data.

On the other hand, implications of the GUP in various fields, such as High Energy Physics, Cosmology and Black Holes have been studied. In \cite{17} potential experimental signatures in some familiar quantum systems are examined. $\b_0$ is a numerical parameter that quantifies the modification strength if we assume that parameter is of the order of unity. This assumption renders quantum gravity effects too small to be measured \cite{17}. But the $\b_0$ dependent terms are important when energies (momenta) are comparable to Planck energy (momenta) and length is comparable to the Planck length. If we cancel this choice, the current experiments predict large upper bounds on it, which are compatible with current observations. It is agreed that such an intermediate length scale, $l_{inter}\simeq l_{Pl}\sqrt{\b_0}$ cannot exceed the electroweak length scale $\sim 10^{17}l_{Pl}$ \cite{17}. This means that $\b_0\leq 10^{34}$. The authors in \cite{17,18} determined a bound in $\b_0$ parameter which proves that GUP could be measured in a low energy system like Landau levels, Lamb shift, potential barrier and  potential step. They calculated the intermediate length scales such that $l_{inter}\simeq 10^{18} l_{Pl},10^{25} l_{Pl}$ and $10^{10} l_{Pl}$ of which the first two are far bigger than the electroweak scale and the last is smaller but may be get further constrained with increased accuracies. In this paper the boundary values of $\b_0$ parameter, in cases of main-sequence stars are $10^{48}, 10^{47}$ and $10^{48}$ and that from white dwarf are $10^{44}$ and $10^{43}$. These can be converted to intermediate length scales, take the average,  $l_{inter} \approx 10^{24} l_{Pl}$ for main-sequence stars and $l_{inter} \approx 10^{22} l_{Pl}$ for white dwarf. They are far bigger than the electroweak scale but it is in approximately compatible with the results addressed by \cite{18} in one case, and it can be considered a reasonable values in the astrophysical regime.

\section{Appendix A}
In general the gravitational field inside the star can be described by a gravitational potential $\f$, which is a solution of the Poisson equation

\begin{equation}
\nabla^2\f=4\pi G\r
\label{p1}
\end{equation}

 for spherical symmetry Poisson equation reduces to

\begin{equation}
\dfrac{1}{r^2}\dfrac{d}{dr}\le(r^2\dfrac{d\f}{dr}\ri)=4\pi G\r
\label{p2}
\end{equation}

In hydrostatic equilibrium, beside Eq.(\ref{p2}) it requires

\begin{equation}
\dfrac{dP}{dr}=-\dfrac{d\f}{dr}\r
\label{p3}
\end{equation}

For a homogeneous gaseous sphere, if we assume that the pressure of the star can be calculated by the modified apolytropic relation

\begin{equation}
P=K_1\r^{\C_1}+\b K_2 \r^{\C_2}
\label{p4}
\end{equation}

The prototype constants $K_1$ and $K_2$ is fixed and can be calculated from nature constants. If $\C_1$ and $\C_2$ are not equal to unity, use Eq.(\ref{p4}) into Eq. (\ref{p3}), and integrate with the boundary conditions $\f=0$ and $\r=0$ at the surface of the sphere, one gets

\begin{equation}
\f=-K_1(1+n_1)\r^{\dfrac{1}{n_1}}-\b K_2(1+n_2)\r^{\dfrac{1}{n_2}}
\label{p5}
\end{equation}

Solving this equation for $\r$, keeping only the terms proportional to $\b$ and use $n_j=\dfrac{1}{\C_j-1}$, then

\begin{equation}
\r=\le[\dfrac{-\f}{K_1(1+n_1)}\ri]^{n_1}-\b\dfrac{n_1 K_2(1+n_2)}{K_1(1+n_1)}\le[\dfrac{-\f}{K_1(1+n_1)}\ri]^{m}
\label{p6}
\end{equation}

where

\begin{equation}
m=\dfrac{n_1}{n_2}+n_1-1
\label{p7}
\end{equation}

Us $\r$ from Eq.(\ref{p6}) into Eq.(\ref{p2}) one gets

\begin{equation}
\dfrac{d^2\f}{dr^2} +\dfrac{2}{r}\dfrac{d\f}{dr}=4\pi G\le[\dfrac{-\f}{K_1(1+n_1)}\ri]^{n_1}-4\pi G\b~\dfrac{n_1 K_2(1+n_2)}{K_1(1+n_1)}\le[\dfrac{-\f}{K_1(1+n_1)}\ri]^{m}
\label{p8}
\end{equation}

Now let us define

\begin{equation}
z=Ar~~~~,~~~~A^2=\dfrac{4\pi G}{K_1^{n_1}(1+n_1)^{n_1}}(-\f_c)^{n_1-1}~~~~,~~~~\o=\dfrac{\f}{\f_c}
\label{p9}
\end{equation}

where $\f_c$ is the potential at the center, it can be defined, like Eq.(\ref{p5}), as

\begin{equation}
\f_c=-K_1(1+n_1)\r_c^{\dfrac{1}{n_1}}-\b K_2(1+n_2)\r_c^{\dfrac{1}{n_2}}
\label{p12}
\end{equation}

Using Eqs.(\ref{p5},\ref{p12}) to define $\o$ as a function of density, one finds that

\begin{equation}
\o=\dfrac{\f}{\f_c}=\le(\dfrac{\r}{\r_c}\ri)^{\dfrac{1}{n_1}}+\b~\dfrac{K_2(1+n_2)}{K_1(1+n_1)}
\le(\dfrac{\r^{n_1}}{\r^{n_2}_c}\ri)^{\dfrac{1}{n_1n_2}}\le[1-\le(\dfrac{\r}{\r_c}\ri)
^{\dfrac{n_2-n_1}{n_1n_2}}\ri]
\label{p13}
\end{equation}

Using Eqs.(\ref{p9}) into Eqs.(\ref{p8}), one gets

\begin{equation}
\dfrac{d^2\o}{dz^2} +\dfrac{2}{z}\dfrac{d\o}{dz}+\o^{n_1}+\eta~ \o^m=0
\label{p10}
\end{equation}

where

\begin{equation}
\eta=\dfrac{\b~ n_1 K_2(1+n_2)}{\f_c}\le(-\f_c\dfrac{A^2}{4\pi G}\ri)^{\dfrac{1}{n_2}}
\label{p11}
\end{equation}

$\eta$ is a dimensionless quantity. Eq.(\ref{p10}) is the modified Lane-Emden equation, We are interested with solutions that are finite at the center, $z=0$. So this equation can be solved with the boundary conditions $\o(0)=1$ and $\o'(0)=1$. $\o$ can be calculated from Eq.(\ref{p10}) and it can be used to determine the mass, density and pressure of the astrophysical objects as a function of a radius.

\end{document}